# Why is the bandwidth of sodium observed to

## in photoemission experime

H. Yasuhara, S. Yoshinaga

*Department of Physics, Graduate School of Sciences, Tohoku Universi*
*Japan*

cond-mat/9905250  18 May 1999

## Abst

The experimentally predicted narrowing in the bandwidth of sodi
the

non-local self-energy effect on quasi-particle energies of the elec

energy correction is a monotonically increasing function of the wave

analysis of photo-emission experiments assumes the final state energ
electron-

like model and hence it incorrectly ascribes the non-local self-ene

energies to the occupied state energies, thus leading to a seeming

PACS numbers: 71.10Ca, 71.20.Dg

Sodium is the most suitable metal for testing the validity of electron liquid at metallic densities since it is almost free from band stru[cture]. angle-resolved photo-emission experiments Plummer and his co-workers possibility that the occupied energy-band of sodium might be remarka[bly narrow in] comparison with the usual value of nearly-free-electron theory. A number of a[uthors have] attempted to interpret the band narrowing of sodium in terms of the occupied band using the GW approxmation or its modifications which a[re] corrections after the manner of Hubbard. The definite answer to this problem ha[s not been obtained in] spite of all these efforts. In this letter we give a clear explanation of wh[y the band is] observed to be narrower in photo-emission experiments, based on the self-energy operator for the electron liquid. We think that the photo-e[mission data of] sodium, if correctly interpreted, may support our assertion that the bandwidth [of sodium is] somewhat broadened under the influence of electron-electron interac[tions].

Energy and momentum conservation in photo-emission experiments [gives]

$$p_f = p_i + G, \qquad h\omega = \varepsilon_f - \varepsilon_i$$

where $h\omega$ is the incoming photon energy; $\varepsilon_i$ and $\varepsilon_f$ are the initial an[d final state energies] respectively; the final state momentum $p_f$ can differ from the initi[al one by] reciprocal lattice vectors G of solids. The occupied energy-band of sodium ha[s been] determined on the basic assumption that the final state energies may [be given by a] free-electron-like model, i. e., $\varepsilon_f = h^2(p+G)^2/2m$ or its simple modificatio[ns in photo-] emission experiments. In this paper we thoroughly consider sodium a[s]

liquid perturbed by a very weak pseudo-potential rather than a non-i
this

case the initial and the final state energies are defined as the qu
uniform

electron liquid and hence the energy conservation relation is given

$$h\omega = E(p+G) - E(p)$$

where E(p) denotes the quasi-particle energy of the electron liquid
electron

energy $\varepsilon_p = h^2 p^2 / 2m$ by the wavenumber-dependent self-energy correctio

We shall begin with a detailed discussion on the wavenumber-de

correction. The self-energy operator can be rigorously expressed a

$$\Sigma_\sigma(p,\varepsilon\,[G]) = \sum_{\sigma'} \int \frac{d\boldsymbol{p}'}{(2\pi)^3} d\varepsilon' \sum_n \frac{1}{2n-1} I^{\sigma\sigma'(n)}(p\varepsilon,\ p'\varepsilon'[G]) G_{\sigma'}(p',\varepsilon') \qquad (3)$$

where $I^{\sigma\sigma'}(p\varepsilon,\ p'\varepsilon'[G])$ denotes the contribution to the so-called par

interaction from all the possible n-th order skeleton diagrams. It

[G])/$\delta G\sigma'$(p', $\varepsilon'$) is the particle-hole irreducible interaction $I^{\sigma\sigma'}$(

above is superior to the standard expression composed from G, W and
highly

symmetric form as a functional of G and hence suitable for including

corrections up to higher orders far beyond the GW approximation. I

difficult to treat $\Sigma$(p, $\varepsilon$ [G]) as a functional of the renormalized G
we

approximately consider the self-energy operator as a functional of t

function $G^0$ and instead take the first iterative solution of Dyson's
[$G^0$]) in

place of the self-consistent solution E(p) [8]. The summation over

expression for $\Sigma_\sigma(p, \varepsilon[G^0])$ by the following coupling-constant-integ

$$\Sigma_\sigma(p,\varepsilon\,[G^0]) = \frac{(e^2)^{\frac{1}{2}}}{2}\int de^2 (e^2)^{-\frac{3}{2}}\sum_{\sigma'}\int\frac{d\mathbf{p}'}{(2\pi)^3}d\varepsilon' I^{\sigma\sigma'}(p\varepsilon, p'\varepsilon'[G^0])G_{\sigma'}(p',\varepsilon') \quad (4)$$

$$I^{\sigma\sigma'}(p\varepsilon,\,p'\varepsilon'[G^0]) = \sum_n I^{\sigma\sigma'(n)}(p\varepsilon,\,p'\varepsilon'[G^0])$$

The replacement of G with $G^0$ in the functional probably will be tole
description

of the coupling between plasmons and multi-pair excitations. Such a
the

accurate evaluation of the self-energy function around $p=p_F$ up to $p=p_F + q_c$ where $q_c$ is
cut-

off wavenumber. For the quantitative evaluation of many-body effec
at

metallic densities it is crucial that the form of $I^{\sigma\sigma'}(p\varepsilon,\,p'\varepsilon'[G^0])$

sufficiently accurate as a functional of $G^0$. From the diagrammatic
that

the irreducible interaction $I^{\sigma\sigma'}(p\varepsilon,\,p'\varepsilon'[G])$ satisfies an integral
diagrammatically

represented in Fig. 5 in Ref. 6. Then, the irreducible interactio
satisfy

the integral equation in order that the resulting self-energy funct:
densities.

Furthermore, the coupling constant integration which is responsible

entering the original equation is indispensable for the qualitative
the

self-energy correction at metallic densities. This implies that the
counted

correctly. For the spin-parallel part of the irreducible interact:
solution

for the integral equation, which is given by eq.(3.13) in Ref. 6, w the spin-antiparallel part of the irreducible interaction we deal with  constant for the moment because it is minor compared with the corre part. The approximate solution for the irreducible interaction $I^{\sigma}$ substantially reducing the screening strength in the GW self-energy calculated self-energy correction in the present theory, i.e., Re $\Sigma$ above is plotted as a function of p for $r_s$=4.0 approximately corresponding comparison the self-energy correction in the GW approximation is als order to sketch out the wavenumber depenendent features of the self- here separate the wavenumber variable into two regions, namely, the region $0<p\leq q_c$ and the off Fermi sphere region where but is not extreme far from $p_F$ In the Slater-approximation-like regions $0 \leq p \leq p_F$ of Re$\Sigma(p, \varepsilon_p$ [G the present theory is rather gradual but positive in contrast with that the off Fermi sphere region, $p > p_F$ the other hand, it approaches much more a function of p. The difference between the two calculations is enla region. In each curve there appears a dip around $q_c$. The dip should disappear in renormalized self-energy functional because plasmons with finite wav widths due to a coupling with multi-pair excitations. The systemat: corrections up to higher orders is thus indispensable for the qualitative and qu self- energy correction not only in the Slater-approximation-like region b sphere region.

In Fig. 2 the energy-band of the electron liquid under the e~ the direction [110]. For comparison the energy-band of the free electro~ figure. The two curves are adjusted so as to agree at the Fermi le~ present theory is somewhat broadened in width and the unoccupied ba~ amount, shifted to the high energy side in comparison with the free ~

In the analysis of photo-emission experiments the final state e~ been conventionally assumed on the nearly-free-electron-like model, leav~ correction to the final states out of account. In other words, the usual assumpt~ has mistaken $\varepsilon_p + \text{Re}\Sigma(p, \varepsilon_p [G^0]) - \text{Re}\Sigma(p+G_{110}, \varepsilon_{p+G}[G^0])$ for $E(p)$ ( $=\varepsilon_p + \text{Re}\Sigma(p, \varepsilon_p [G^0])$ is apparent from the figure, the difference $\text{Re}\Sigma(p+G_{110}, \varepsilon_{p+G_{110}}[G^0]) - \text{Re}\Sigma(G_{110}, \varepsilon_{G_{110}}[G^0])$ much larger than the difference $\text{Re}\Sigma(p_F, \varepsilon_{p_F}[G^0]) - \text{Re}\Sigma(0, 0 [G^0])$. From these ~ we conclude that the narrowing in the occupied bandwidth of sodium is a~ assumption of the final state energies. Note that the occupied ban~ somewhat broader if as the final state energies one adopts the quas~ in place of the free electron energies $\varepsilon_{p+G_{110}}$.

In order to make a direct comparison with experiment possible ~ virtual energy-band with the dispersion $E^{\text{vir}}(p) = \varepsilon_p + \text{Re}\Sigma(p, \varepsilon_p [G^0]) - \text{Re}\Sigma(p+G_{110}, ~$ which is

the theoretical analogue of the experimental energy-band obtained f~

An excellent agreement between the experimental band [2] and the calcul[ated band is] seen from Fig.3. Not only the magnitude of the narrowing in the bandwit[h but also the] features of the experimental energy-band are almost completely interpreted i[n terms of the self-] energy effect on the quasi-particle energies of the electron liquid[. The dip] around $p=0.2p_F$ in the virtual energy-band curve is ascribed to the fac[t that] $\text{Re}\Sigma(p,\varepsilon_p[G^0]) - \text{Re}\Sigma(p+G_{110},\varepsilon_{p+G_{110}}[G^0])$ is an almost linearly decreasing function o[f] wavenumbers.

According to the present theory, the slope of the self-energy c[urve is] gradually steepened with increasing $r_s$ in the off Fermi sphere regio[n.]

over the entire range of metallic densities in the Slater-approxima[tion. Thus] the calculated virtual bandwidth correspondiong to the experimental band[, at the] density level of sodium, is much less narrowed for $r_s=2.0$ and on the [contrary more] narrowed for $r_s=5.0$. This is consistent with the experimental results of al[uminum and] potassium ($r_s=4.87$) [10], though the quasi-particle energies of thes[e metals are affected] by the crystal potential than those of sodium.

In conclusion, for the quantitative determination of the occup[ied bandwidth of] metals the usual assumption that the final state energies may be identifie[d with free-electron-] like model requires reconsideration from a many-body theoretic point of v[iew because] of course applies to transition metals like nickel. In order to inte[rpret] experiments of metals one must then rely on the Dyson's equation whi[ch includes] non-local self-energy effect on quasi-particle states since the Koh[n-Sham] construction cannot describe such a non-local effect on both the in[itial and final states]. highly desirable that the final state energies be absolutely identified by [inverse photoemission] combined with photoelectron spectroscopy[11].

One of the authors (H.Y.) would like to thank Professors S. Sut[o ... for] many helpful and stimulating discussions. He is also thankful to Dr K. [... for] conversations.

Figure Captions

Fig. 1.

The self-energy correction in the present theory is plotted as a fur
comparison the correction in the GW approximation is also plotted.

Fig. 2.

The energy-band of the electron liquid under the empty lattice poter
[110]; bold line: present theory, thin line: free electron theory.

Fig. 3.

A comparison between the experimental energy-band and the calculated
line: virtial energy-band in the present theory, dotted line: free triangles: experimental results in Ref. 2.